\documentclass{ws-ijmpcs}

\begin{document}

\def\bea{\begin{eqnarray}}
\def\eea{\end{eqnarray}}
\def\beq{\begin{equation}}
\def\eeq{\end{equation}}

\markboth{Kiwoon Choi}
{Thermal  Production OF Axino Dark  Matter}

%
\catchline{}{}{}{}{}
%

\title{THERMAL PRODUCTION OF AXINO DARK MATTER
}

\author{KIWOON CHOI}

\address{Physics Department, Korea Advanced Institute of Science and Technology \\
Daejeon 305-701, South Korea
\\
kchoi@kaist.ac.kr}

\maketitle

\begin{history}
\received{Day Month Year}
\revised{Day Month Year}
\end{history}

\begin{abstract}
We discuss certain features of the low energy effective interactions of axion
supermultiplet, which are relevant for axino cosmology, and examine 
the implication to
thermal production of axino in the early Universe.

\end{abstract}

\ccode{PACS numbers:}

\section{Introduction}	

Supersymmetric and axionic extension of the standard model provides an
appealing solution to both of  
the gauge hierarchy problem and the strong CP problem.
In such model, the superpartners of axion, i.e. the axino and saxion,  can
have a variety of cosmological implications\cite{axino,axi_cos}. In particular,
 axino can be a good candidate for cold dark matter, depending upon 
 the mechanism of
supersymmetry breaking and cosmological evolution in the early Universe.
Even when axino is not stable, so does not constitute the dark matter, it
can affect the 
evolution of early Universe in various ways. For instance, late
decays of axino might affect the
relic dark matter density  and/or the Big-Bang nucleosynthesis and/or the 
large scale structure
formation\cite{axino_decay}.

One of the  key issues in axino cosmology  is the
thermal production of axino  by scattering or decay of particles in
thermal equilibrium  in the early Universe\cite{axino_thermal}. Most of the previous
analysis of thermal axino production\cite{axino_thermal} is based on the
 local effective interaction of the form \begin{equation}
 \int
d^2\theta \, \frac{1}{32\pi^2}
\frac{A}{v_{PQ}}W^{a\alpha}W^a_\alpha,\label{eff1}\end{equation}
 where $v_{PQ}$ is the
scale of spontaneous breakdown of the PQ symmetry,
and 
$A=(s+ia)/\sqrt{2}+\sqrt{2}\theta \tilde a +\theta^2 F^A$ is the axion
superfield which contains the axion $a$, the saxion $s$, and the
axino $\tilde a$ as its component fields.
In some cases, for instance the KSVZ-type  model with
heavy quark supermultiplet  having a mass $M_Q\sim v_{PQ}$, the effective interaction
(\ref{eff1}) provides a good description of the low energy 
dynamics of axion supermultiplet.
However, in other cases, e.g. the KSVZ-type model with
$M_Q\ll v_{PQ}$ or the DFSZ-type model without exotic heavy quark, analysis using the effective interaction
(\ref{eff1}) alone 
yields a highly overestimated axino production rate
as the correct rate experiences a cancellation due to 
other effective interactions\cite{bae}.
In this talk\footnote{This talk is based on Ref. 5.}, we discuss first generic structure
of the low energy effective interactions of axion
supermultiplet in models having a UV completion 
in which the PQ symmetry is linearly realized,  and
then consider its implication
to cosmological axino production.

\section{Effective interactions  of axion supermultiplet}

Generic Wilsonian effective lagrangian of the axion
superfield at energy scale $\Lambda$  below the PQ scale $v_{PQ}$ 
takes the form \bea \label{eff_general} {\cal L}_{\rm eff}(\Lambda)  &=& \int
d^4\theta \, \left( K_A(A+A^\dagger)+
Z_n(A+A^\dagger)\Phi_n^\dagger\Phi_n\right)\nonumber \\
 &+&\left[\, \int d^2\theta \, \left( \frac{1}{4} f^{\rm
eff}_a(A)W^{a\alpha}W^a_\alpha+W_{\rm eff}\right)+{\rm
h.c}\,\right],\eea where $\{\Phi_n\}$ denote the light gauge-charged
matter fields, and
 \bea \label{eff_general1} K_A &=& \frac{1}{2}(A+A^\dagger)^2
+{\cal O}\left(\frac{(A+A^\dagger)^3}{v_{PQ}}\right),
\nonumber \\
\ln Z_n&=& \left.\ln Z_n\right|_{A=0}+\tilde
y_n\frac{(A+A^\dagger)}{v_{PQ}}+{\cal
O}\left(\frac{(A+A^\dagger)^2}{v^2_{PQ}}\right),\nonumber \\
 f^{\rm eff}_a &=&
 \frac{1}{\hat g_a^2(\Lambda)}-\frac{C^a_W}{8\pi^2}\frac{A}{v_{PQ}},
\nonumber
\\
W_{\rm eff}&=& \frac{1}{2} e^{-(\tilde x_n+ \tilde
x_m)A/v_{PQ}}M_{mn}\Phi_m\Phi_n \nonumber \\
& +& \frac{1}{6}e^{-(\tilde x_n+\tilde
x_m+\tilde x_p)A/v_{PQ}}\lambda_{mnp}\Phi_m\Phi_n\Phi_p. \nonumber \eea The PQ
symmetry is realized as 
$U(1)_{PQ}: A\rightarrow A+i\alpha v_{PQ},
\Phi_n\rightarrow e^{i \tilde x_n\alpha}\Phi_n$, and the
Wilsonian couplings between the axion superfield  and the gauge/matter 
superfields are given
by
 \bea
 \label{i1}
\Delta_1 {\cal L}(\Lambda) &=& -\int d^2\theta \,
\frac{C^a_W}{32\pi^2}\frac{A}{v_{PQ}}W^{a\alpha}W^a_\alpha, \nonumber
 \\
 \label{i2}
\Delta_2 {\cal L}(\Lambda)&=&\int d^4\theta \,\, \tilde
y_n\frac{(A+A^\dagger)}{v_{PQ}} \Phi_n^\dagger \Phi_n,
\nonumber  \\
\label{i3}\Delta_3 {\cal L}(\Lambda)&=&-\int d^2\theta
\,\frac{A}{v_{PQ}} \Big[\, \frac{(\tilde x_m+\tilde x_n)}{2}M_{mn}
\Phi_n\Phi_m \nonumber \\
&& \qquad +\, \frac{(\tilde x_m+\tilde x_n+\tilde
x_p)\lambda_{mnp}}{6}\Phi_m\Phi_n\Phi_p\,\Big]. \label{eff_coupling}\eea
Then there are three quantities 
$\{\, C_W^a,\, C^a_{PQ},\, C^a_{1PI}\,\}$
which are related to the
axino coupling to gauge supermultiplets,
where $C_W^a$ are the Wilsonian couplings  in (\ref{i1}), $C^a_{PQ}$
are the PQ anomaly coefficients defined as
 \bea
\partial_\mu
J^\mu_{PQ}=\frac{g^2}{16\pi^2} C_{PQ}^a F^{a\mu\nu}\tilde
F^a_{\mu\nu},\eea
 and finally $C_{1PI}^a$ determines the leading part of the 1PI axino-gaugino-gauge boson amplitude\bea {\cal
A}^a_{1PI}(k,q,p) = -\frac{g^2}{16\pi^2 \sqrt{2}v_{PQ}}\tilde
C^a_{1PI}\delta^4(k+q+p) \bar{u}(k)\sigma_{\mu\nu} \gamma_5
v(q)\epsilon^\mu p^\nu \eea which shows the behavior \bea
\label{1pi_def} && \tilde C_{1PI}^a(k^2=q^2=0, \,M^2_{\rm light}< p^2 < M^2_{\rm heavy})\nonumber \\
& = & C_{1PI}^a+{\cal O}\left(\frac{M_{\rm
light}^2}{p^2}\ln^2\left(\frac{p^2}{M_{\rm light}^2}\right)\right)
+{\cal O}\left(\frac{p^2}{M_{\rm heavy}^2}\right) , \eea where
$M_{\rm light}$ and $M_{\rm heavy}$
 denote the masses of matter fields in the effective theory
(\ref{eff_general}).
 It is then straightforward to find\cite{bae} \bea \label {pqanomaly}&& C^a_{PQ} \, =\,
C^a_W + 2\sum_n \tilde x_n {\rm
Tr}(T_a^2(\Phi_n)), \nonumber  \\
\label{c_1pi} && 
C_{1PI}^a(M_\Phi^2 <p^2<\Lambda^2)\,=\,\frac{C^a_W(\Lambda)-2\sum_n \tilde
y_n(p){\rm Tr}(T_a^2(\Phi_n))}{1-{\rm
Tr}(T_a^2(G))g_a^2(p)/8\pi^2}, \nonumber
 \eea where
$M_{\Phi}$ is the mass of the heaviest PQ-charged and gauge-charged
matter field in the model, and
$\tilde y_n(p ) =
v_{PQ}\left.{\partial \ln {\cal Z}_n}/{\partial
A}\right|_{A=0}$
for the 1PI
wavefunction coefficient ${\cal Z}_n$ of $\Phi_n$,
which can be chosen to satisfy
the matching condition ${\cal
Z}_n(p^2=\Lambda^2)=Z_n(\Lambda)$.

Within the effective theory (\ref{eff_general}), one can make a
holomorphic field redefinition  $\Phi_n
\rightarrow e^{z_n A/v_{PQ}} \Phi_n$, after which the PQ symmetry is given by  $U(1)_{PQ}: A\rightarrow A+i\alpha v_{PQ}, 
\Phi_n\rightarrow e^{i(\tilde x_n-z_n)\alpha}\Phi_n$, and the
Wilsonian couplings of the axion superfield are changed as
 \bea
 \label{change}
 &&  C^a_W
\rightarrow C^a_W +2\sum_n z_n{\rm
Tr}(T_a^2(\Phi_n)),\nonumber \\
&& \tilde y_n \rightarrow \tilde y_n +z_n,\quad \tilde x_n
\rightarrow \tilde x_n-z_n.\eea  Note that
$C_{PQ}^a$ and $C_{1PI}^a$ are directly linked to observables, and
therefore invariant under the reparametrization (\ref{change}) of
the Wilsonian couplings.

A key result of our discussion, which has direct implication for
cosmological axino production, is that the 1PI axino-gaugino-gauge
boson amplitude in the momentum range $M_\Phi^2 < p^2 < v_{PQ}^2$ is
suppressed by $M_\Phi^2/p^2$, more specifically\cite{bae}
 \bea
\label{1pi_zero} \tilde C_{1PI}^a(M_\Phi^2 < p^2
<v^2_{PQ})
={\cal
O}\left(\frac{M_\Phi^2}{p^2}\ln^2\left(\frac{M_\Phi^2}{p^2}\right)\right),\eea
As we will see  below, the result (\ref{1pi_zero}) applies to
generic supersymmetric axion model if the model has a UV realization
at $M_*\gg v_{PQ}$, in which (i) the PQ symmetry is linearly realized  in the
standard manner, i.e. $U(1)_{PQ}:  \Phi_I
\rightarrow e^{ix_I\alpha}\Phi_I$, where $\{\Phi_I\}$ stand for
generic chiral matter superfields, and (ii) all higher dimensional
operators of the model are suppressed by appropriate powers of
$1/M_*$.

To see this, let $\{\Phi_A\}$ denote the gauge-singlet but
generically PQ-charged matter fields, whose VEVs break $U(1)_{PQ}$
spontaneously,  and $\{\Phi_n\}$ denote the gauge-charged matter
fields in the model.
 Then the K\"ahler
potential and superpotential at the UV scale $M_*$ can be expanded
in powers of the gauge-charged matter fields as follows \bea
K&=&K_{PQ}(\Phi_A^\dagger,\Phi_A)+\Big(1+ \frac{\kappa_{\bar AB
n}}{M_*^2}\Phi_A^\dagger \Phi_B+...\Big)
\Phi_n^\dagger \Phi_n+..., \nonumber \\
W&=& W_{PQ}(\Phi_A) +
\frac{1}{2}\Big(\hat\lambda_{Amn}\Phi_A+\frac{\hat
\lambda_{ABmn}}{M_*}\Phi_A\Phi_B+...\Big)\Phi_m\Phi_n\nonumber \\
&&+ \, \frac{1}{6}\Big(\hat\lambda_{mnp}+
\frac{\hat\lambda_{Amnp}}{M_*}\Phi_A+...\Big)\Phi_m\Phi_n\Phi_p+...,
\label{full_th}\eea
where $K_{PQ}$ and $W_{PQ}$ are the K\'ahler potential and
superpotential of the PQ sector fields $\{\Phi_A\}$,
$M_*$ is  presumed to be the Planck scale or the GUT
scale,
 and the
ellipses stand for higher dimensional terms. 
Under the assumption that $K_{PQ}$ and $W_{PQ}$ provide a proper
dynamics to break the PQ symmetry spontaneously, we can parameterize
the PQ sector fields as $\Phi_A =
\left(\frac{v_A}{\sqrt{2}}+ U_{Ai}\rho_i\right)e^{x_A
A/v_{PQ}}$, where $v_A =\langle \Phi_A\rangle$ with
$v_{PQ}^2=\sum_A x_A^2 |v_A|^2$, $\rho_i$ denote the massive chiral
superfields in the PQ sector, and $U_{Ai}$ are the mixing
coefficients which are generically of order unity. For this
parametrization, the K\"ahler potential and
superpotential at $M_*$ take the form \bea K &=&
K_{PQ}(\rho_i^\dagger, \rho_i, A+A^\dagger) + \left(Z_n^{(0)}+{\cal
O}\Big(\frac{v_{PQ}}{M_*^2}(A+A^\dagger),\frac{v_{PQ}\rho_i}{M_*^2},\frac{v_{PQ}\rho^\dagger_i}{M_*^2}\Big)\right)
 \Phi_n^\dagger \Phi_n+...,
\nonumber \\
W &=& W_{PQ}(\rho_i) +\frac{1}{2} \left(M_{mn}+{\cal
O}\Big(\frac{M_{mn}\rho_i}{v_{PQ}}\Big)\right)e^{-(x_m+x_n)A/v_{pQ}}\Phi_m\Phi_n\nonumber
\\
&&+\, \frac{1}{6}\left(\lambda_{mnp}+{\cal
O}\Big(\frac{\rho_i}{M_*}\Big)\right)e^{-(x_m+x_n+x_p)A/v_{PQ}}\Phi_m\Phi_n\Phi_p+...,\eea
where $Z_n^{(0)}$ and $M_{mn}$ are field-independent constants, and
 the Yukawa coupling constants $\lambda_{mnp}$ obeys the PQ selection rule \bea
 \label{pq_sel}
(x_m+x_n+x_p)\lambda_{mnp}=(x_m+x_n+x_p)\left(\hat\lambda_{mnp}+{\cal
O}\left(\frac{v_{PQ}}{M_*}\right)\right)={\cal
O}\left(\frac{v_{PQ}}{M_*}\right).\eea
One can now  integrate out the massive $\rho_i$ as well as the high
momentum modes of light fields, and also make an arbitrary  field redefinition
$\Phi_n\rightarrow e^{z_nA/v_{PQ}}\Phi_n$ to derive an effective theory in generic field basis.
The resulting effective lagrangian at $\Lambda$ just below
$v_{PQ}$ takes the form of (\ref{eff_coupling}) with \bea
\label{estimate} C^a_W &=& -8\pi^2 v_{PQ}\frac{\partial f_a^{\rm
eff}}{\partial A}= 2\sum_n z_n {\rm
Tr}(T_a^2(\Phi_n)), \nonumber \\
 \tilde x_n &=& x_n
-z_n,
\nonumber \\
\tilde y_n(\Lambda) &=& v_{PQ}\left.\frac{\partial \ln Z_n}{\partial
A}\right|_{A=0}= 
\,=\, z_n + {\cal O}\left(\frac{M_{\Phi}^2}{v_{PQ}^2},\frac{v_{PQ}^2}{M_*^2}\right).\eea 
 We
then have \bea \label{boun} C_{1PI}^a(p^2=\Lambda^2 ) =
C_{W}^a(\Lambda)-2\sum_n \tilde y_n(\Lambda){\rm Tr}(T_a^2(\Phi_n))
= {\cal O}\left(\frac{M_\Phi^2}{v_{PQ}^2},\frac{v_{PQ}^2}{M_*^2}\right),\label{boun} \eea  and the PQ selection rule
(\ref{pq_sel}) takes the form \bea (\tilde x_m +\tilde x_n +\tilde
x_p +\tilde y_m +\tilde y_n +\tilde y_p)\lambda_{mnp}= {\cal
O}\left(\frac{M_{\Phi}^2}{v_{PQ}^2},\frac{v_{PQ}}{M_*}\right).\nonumber \eea
Including the 1PI RG evolution, the above estimate is valid for external momentum  in the range $M_\Phi
< p< v_{PQ}$, so the estimate (\ref{1pi_zero})
 is valid even when higher loop effects are
taken into account.
This is in fact a simple consequence of
that the axion supermultiplet is decoupled from gauge and matter
supermultiplets  in
the limit $M_\Phi\rightarrow 0$ and $M_*\rightarrow \infty$, which is manifest in the full theory
(\ref{full_th}).
With the boundary condition (\ref{boun}), one can
determine $C_{1PI}^a$ at lower momentum scale $p<M_\Phi$ by
computing the threshold correction,  which yields  \bea C_{1PI}^a(p)
&=& C_W^a(\Lambda)-2\sum_{M^2_n < p^2} \tilde y_n(\Lambda) {\rm
Tr}(T_a^2(\Phi_n)) +2\sum_{M_m^2>p^2}\tilde x_m(\Lambda) {\rm
Tr}(T_a^2(\Phi_m)) \nonumber \\
&=& 2\sum_{M^2_m>p^2} \Big(\tilde x_m(\Lambda)+\tilde
y_m(\Lambda)\Big) {\rm Tr}(T_a^2(\Phi_m)), \eea where $C_{W}^a(\Lambda),
\tilde y_n(\Lambda)$ and $\tilde x_n(\Lambda)$ are the Wilsonian
couplings in the effective lagrangian (\ref{eff_general}) at the
cutoff scale $\Lambda$ just below $v_{PQ}$.

\section{Thermal production of axino}

To discuss thermal axino production,
we choose a field basis in which the Wilsonian  couplings
of axion supermultiplet at $\Lambda$  are given by \bea
\label{basis1}C_W(\Lambda)&=& 0, \quad \tilde x_n \, = \,x_n, \quad
\tilde y_n(\Lambda)\,=\, {\cal O}\left(\frac{M_\Phi^2}{v_{PQ}^2},
\,\frac{v_{PQ}^2}{M_*^2}\right),\eea
which is always possible under the boundary condition (\ref{boun}),
and convenient for describing the
physics at energy scales  in the range $M_\Phi< E< v_{PQ}$,
since the decoupling of the
axion supermultiplet in the limit $M_\Phi\rightarrow 0$ is manifest.

Let $\Phi,\Phi^c$ denote the heaviest PQ-charged and gauge-charged
matter superfield with a supersymmetric mass $M_\Phi$. In the field
basis (\ref{basis1}),  the relevant effective interaction of axion
supermultiplet takes a simple form \bea \label{eff} -\int d^2\theta
\, (x_\Phi+x_{\Phi^c}) M_\Phi\frac{A}{v_{PQ}} \Phi\Phi^c+{\rm
h.c},\eea where we have ignored the small $\tilde y_n={\cal
O}(M_\Phi^2/v_{PQ}^2, v_{PQ}^2/M_*^2)$.
 A key element for the
axino production by gauge supermultiplet is the 1PI axino-gaugino-gauge boson
amplitudes which are given by
\begin{eqnarray}
\widetilde{C}_{\text{1PI}}(k^2=q^2=0; \, p^2\gg M_\Phi^2)
&\simeq&(x_\Phi+x_{\Phi^c})\frac{M_\Phi^2}{p^2}\ln^2\biggl(\frac{p^2}{M_\Phi^2}\biggr)
\nonumber \\
\widetilde{C}_{\text{1PI}}(k^2=q^2=0; \, p^2\ll M_\Phi^2)
&=&x_\Phi+x_{\Phi^c}+{\cal O}\biggl(\frac{p^2}{M_\Phi^2}\biggr).
\label{1PI_coupling}
\end{eqnarray}
With these 1PI amplitudes and also the axino-matter coupling
(\ref{eff}), we can calculate the thermal production of axino in
the temperature range of our interest.  

To proceed, let us  consider the axino production processes $I+J\rightarrow \tilde a + K$, where $I,J,K$ stand for the particles 
in gauge or matter supermultiplets.
The 1PI amplitudes (\ref{1PI_coupling}) imply that the amplitude
of the axino production through the transition $g\rightarrow \tilde
g +\tilde a$ in the temperature range $M_\Phi\ll T < v_{PQ}$ is
suppressed by $M_\Phi^2/T^2$. As a result, in this temperature
range, axinos are produced mostly by the transition $\Phi\rightarrow
\tilde \Phi +\tilde a$ or $\tilde \Phi\rightarrow \Phi+\tilde a$,
 and  the production
rate is given by \bea \Gamma_{\tilde a}(M_\Phi\ll T<v_{PQ}) 
= {\cal O}(1)\times\frac{g^2M_\Phi^2T^4}{\pi^5 v_{PQ}^2}.
 \eea  On the other hand, at lower temperature
$T\ll M_\Phi$, the matter multiplet $\Phi$ is not available anymore,
and axinos are produced mostly by the
transition $g\rightarrow \tilde g +\tilde a$ or $\tilde
g\rightarrow g+\tilde a$, which results
in
\begin{equation}
\Gamma_{\tilde a}( T \ll M_\Phi) ={\cal O}(1) \times
\frac{g^6T^6}{64\pi^7v_{PQ}^2}.
\end{equation}
Solving the Boltzmann equation, the relic axino number density over
the entropy density can be determined as 
\begin{equation}
Y_{\tilde a}\equiv \frac{n_{\tilde a}(T_0)}{s(T_0)}=\int_{T_0}^{T_R}
\frac{dT}{T} \frac{\Gamma_{\tilde a}}{s(T)H(T)}
\end{equation}
where $T_R$ is the reheat temperature, $s(T)=2\pi^2g_*T^3/45$ is the
entropy density, and $H(T)=\sqrt{\pi^2g_*/90}T^2/M_{Pl}$ is the
Hubble parameter for the effective degrees of freedom $g_*$ and the
reduced Planck mass $M_{Pl}=2.4\times10^{18}$ GeV. We then find \bea
Y_{\tilde a}( T_R \ll M_\Phi) &=& {\cal O}(1)\frac{\bar gg^6M_{Pl}}{64\pi^7
v_{PQ}^2}T_R,\nonumber \\
Y_{\tilde a}(M_{\Phi}\ll T_R \ll v_{PQ}) &=& {\cal O}(1)
\frac{\bar gg^2M_{Pl}}{2\pi^4v_{PQ}^2}M_\Phi,
\label{relic2}\eea where $\bar g ={135\sqrt{10}}/{2\pi^3g_*^{3/2}}$.

\begin{figure}
\begin{center}
\includegraphics[width=8cm]{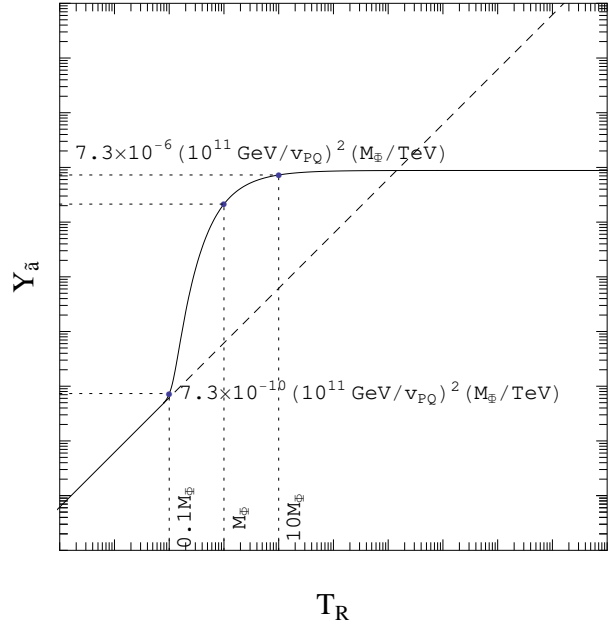}
\end{center}
\caption{Relic axino number density over the entropy density  vs the
reheat temperature $T_R$ (solid line). The dashed line is
the result one would get by using only the effective interaction (\ref{eff1}). 
\label{result_general}}
\end{figure}

Fig. \ref{result_general} summarizes the results of our analysis.
 It shows $Y_{\tilde a}\propto T_R$ for $T_R\lesssim
0.1 M_\Phi$, which is due to that 
axinos are produced mostly
through the transition $g\rightarrow \tilde g+\tilde a$ or $\tilde
g\rightarrow g+\tilde a$ when $T\lesssim 0.1 M_\Phi$.
 If one uses only the
effective interaction (\ref{eff1}) to evaluate the axino production by
$g\rightarrow \tilde g+\tilde a$ or $\tilde g\rightarrow g+\tilde
a$, as one did in most of the previous analysis, one would get
$Y_{\tilde a}\propto T_R$ even for $T_R\gtrsim 0.1M_\Phi$, as
represented by the dashed line in Fig. \ref{result_general}. 
Taking it into account that 
the axino production at $T>
M_\Phi$ is mostly due to the transition $\Phi\rightarrow \tilde
\Phi+\tilde a$ or $\tilde \Phi\rightarrow \Phi+\tilde a$,
one can easily understand the behavior of $Y_{\tilde a}$ for $T_R>10
M_\Phi$, which is nearly  independent of $T_R$. Note that the dashed
line crosses the correct solid line at $T_R\sim 10^3 M_\Phi$,
implying that the previous analysis based on the effective
interaction (\ref{eff1}) alone
gives rise to an overestimated axion relic density for the reheat
temperature $T_R \gtrsim 10^3 M_\Phi$, while it gives an
underestimated $Y_{\tilde a}$ for $0.1 M_\Phi\lesssim T_R\lesssim
10^3M_\Phi$.

\section{Conclusion}

 For supersymmetric axion models  which
 have a UV completion with linearly realized PQ symmetry  
 at a fundamental scale
$M_*\gg v_{PQ}$,
the axion supermultiplet is
decoupled from the gauge and matter supermultiplets in the limit
$M_\Phi/v_{PQ}\rightarrow 0$ and $v_{PQ}/M_*\rightarrow 0$, where
$M_\Phi$ is the mass of the heaviest PQ-charged and gauge-charged
matter multiplet in the model.  As a result, in models with small
values of $M_\Phi/v_{PQ}$ and $v_{PQ}/M_*$, the axino production
rate at temperature $T\gg M_\Phi$ is suppressed by the powers of
small $M_\Phi/T$. 
This feature  is particularly important for the cosmology of
supersymmetric DFSZ axion model\cite{bae,bae1} in which $M_\Phi$ corresponds to the
MSSM Higgs $\mu$-parameter, so is far below $v_{PQ}$.
Cosmology of KSVZ axion model can be significantly altered also, if
the PQ-charged exotic quark has a mass well below $v_{PQ}$.
One immediate consequence (see Fig. \ref{result_general}) is the relic axino density vs the reheat temperature
for $0.1 M_\Phi <T_R < v_{PQ}$, which is quite different  from 
the previous result obtained using the effective interaction (\ref{eff1}) alone.

\section*{Acknowledgement}

This work is supported by the KRF Grants funded by the Korean
Government (KRF-2008-314-C00064 and KRF-2007-341-C00010) and the
KOSEF Grant funded by the Korean Government (No. 2009-0080844).



\begin{thebibliography}{0}    

\bibitem{axino} For recent reviews, see
L.~Covi and J.~E.~Kim,
  New J.\ Phys.\  {\bf 11}, 105003 (2009);
F.~D.~Steffen,
  Eur.\ Phys.\ J.\  C {\bf 59}, 557 (2009).


\bibitem{axi_cos}
See for instance K.~Choi, E.~J.~Chun, J.~E.~Kim,
  Phys.\ Lett.\  {\bf B403}, 209 (1997); K.~Choi, K.~S.~Jeong, W.~-I.~Park and C.~S.~Shin,
  JCAP {\bf 0911}, 018 (2009);
  J.~Fan, M.~Reece and L.~-T.~Wang,
  JHEP {\bf 1109}, 126 (2011).

\bibitem{axino_decay}
E.~J.~Chun, H.~B.~Kim, J.~E.~Kim,
  Phys.\ Rev.\ Lett.\  {\bf 72}, 1956 (1994); 
C.~Cheung, G.~Elor and L.~J.~Hall,
  Phys.\ Rev.\ D {\bf 85}, 015008 (2012); 
A.~Freitas, F.~D.~Steffen, N.~Tajuddin and D.~Wyler,
  JHEP {\bf 1106}, 036 (2011).  
  
  

\bibitem{axino_thermal} 
L.~Covi, H.~B.~Kim, J.~E.~Kim and L.~Roszkowski,
  JHEP {\bf 0105}, 033 (2001);
L.~Covi, L.~Roszkowski and M.~Small,
  JHEP {\bf 0207}, 023 (2002);  
   A.~Brandenburg, F.~D.~Steffen,
  JCAP {\bf 0408}, 008 (2004); A.~Strumia,
  JHEP {\bf 1006}, 036 (2010).
  



\bibitem{bae}
 K.~J.~Bae, K.~Choi and S.~H.~Im,
  JHEP {\bf 1108}, 065 (2011).
    

\bibitem{bae1}
  K.~J.~Bae, E.~J.~Chun and S.~H.~Im,
  arXiv:1111.5962 [hep-ph].
  
 



\end{thebibliography}
\end{document}